\documentclass[journal=jpclcd,manuscript=letter, layout=twocolumn]{achemso}
\setkeys{acs}{doi = true}

\usepackage[fontsize=10pt]{fontsize}

\usepackage{chemformula} 
\usepackage[T1]{fontenc} 

\usepackage[colorlinks]{hyperref}

\usepackage[version=4]{mhchem}

\usepackage{subcaption} 
\usepackage{float}
\usepackage{multirow}
\usepackage{longtable}
\usepackage{ulem}
\usepackage{natbib}
\usepackage{amssymb}
\usepackage{amsmath}
\usepackage{lineno}
\usepackage{adjustbox}
\usepackage{booktabs}
\usepackage{url}
\usepackage{upgreek}

\newcommand{\doi}[1]{\href{http://dx.doi.org/#1}{\nolinkurl{#1}}}

\author{Gabi Wenzel}
\affiliation[mit]
{Department of Chemistry, Massachusetts Institute of Technology, 77 Massachusetts Ave, Cambridge, MA 02139, USA}
\alsoaffiliation[cfa]{Center for Astrophysics | Harvard~\&~Smithsonian, 60 Garden St, Cambridge, MA 02138, USA}
\altaffiliation{Contributed equally to this work}
\author{Miguel Jim\'enez-Redondo}
\affiliation[mpe]
{Max Planck Institute for Extraterrestrial Physics, Giessenbachstrasse 1, 85748 Garching, Germany}
\altaffiliation{Contributed equally to this work}
\author{Milan On\v{c}\'{a}k}
\affiliation[Inn]
{Institute for Ion and Applied Physics, University of Innsbruck, Technikerstraße 25, Innsbruck 6020, Austria}
\author{Brett A. McGuire}
\affiliation[mit]
{Department of Chemistry, Massachusetts Institute of Technology, 77 Massachusetts Ave, Cambridge, MA 02139, USA}
\alsoaffiliation[nrao]
{National Radio Astronomy Observatory, 520 Edgemont Rd, Charlottesville, VA 22903, USA}
\author{Sandra Br\"{u}nken}
\affiliation[felix]
{Radboud University, Institute for Molecules and Materials, FELIX Laboratory, Toernooiveld 7, 6525ED Nijmegen, The Netherlands}
\author{Paola Caselli}
\affiliation[mpe]
{Max Planck Institute for Extraterrestrial Physics, Giessenbachstrasse 1, 85748 Garching, Germany}
\author{Pavol Jusko}
\email{pjusko@mpe.mpg.de}
\affiliation[mpe]
{Max Planck Institute for Extraterrestrial Physics, Giessenbachstrasse 1, 85748 Garching, Germany}

\title{Infrared Spectroscopy of Pentagon-Containing PAHs:\\ Indenyl and Fluorenyl Anions and Indenyl Cation}

\abbreviations{IR,IRPD,IRMPD}

\newcommand{\ie}{i.\,e.}
\newcommand{\rcm}{\ensuremath{\text{cm}^{-1}}}

\newcommand{\kJmol}{\ensuremath{\text{kJ}\,\text{mol}^{-1}}}
\newcommand{\massU}{\ensuremath{m/z}}

\begin{document}





\begin{abstract}
Polycyclic aromatic hydrocarbon (PAH) ions are crucial intermediates in interstellar chemistry and may play a key role in the infrared emission features observed in space. Here, we investigate the infrared spectra of the indenyl (\ce{C9H7-}) and fluorenyl (\ce{C13H9-}) anions and the indenyl cation (\ce{C9H7+}) using infrared pre-dissociation (IRPD) spectroscopy. The experiments were performed in a cryogenic 22 pole ion trap at the FELion beamline of the tunable free-electron laser FELIX. Spectral analysis of the two anionic PAHs, in combination with density functional theory (DFT) computations, revealed key vibrational modes near $1300\,\rcm$, making these ions potential carriers of the $7.7\,\upmu\text{m}$ PAH emission band seen in many astronomical objects. The feature-rich spectrum of cationic indenyl could not be entirely explained by modeling through time-independent anharmonic DFT calculations. Although a better match has been achieved through molecular dynamics simulations, we cannot completely rule out the presence of multiple cationic isomers of the \ce{H}-loss fragments of indene in the experiments.

\end{abstract}


\twocolumn
The physics and chemistry of the interstellar medium (ISM) are governed by molecular processes, involving organic matter, that are intimately linked to the different stages of the stellar evolution cycle. Approximately 10--20$\,\%$ of the cosmic carbon content is locked up in polycyclic aromatic hydrocarbons (PAHs)~\cite{tielens2008} and their presence in the ISM has long been proposed due to the observations of the Aromatic Infrared Bands (AIBs) in many astronomical objects~\cite{peeters2002}. In these regions, isolated PAHs are photo-excited by ultraviolet (UV) photons. One of their major relaxation channels is vibrational relaxation resulting in the emission of infrared (IR) photons that give rise to the AIBs~\cite{joblin2020}. AIB spectra have been used to study the bulk composition of astrophysical PAHs, their structure, functionalization, and charge state~\cite{peeters2011}. The identification of a specific PAH from AIB spectra, however, remains elusive due to the broad emission features resulting from overlapping bands of many different PAHs. The cold and dark Taurus Molecular Cloud 1 (\mbox{TMC-1}, $T \approx 10\,\mathrm{K}$) has proven to be a molecule-rich environment and to date nine PAHs have been unambiguously detected in radio astronomical observations of this source. The majority of these consists of nitrile-functionalized PAHs due to their increased permanent electric dipole moments, namely 2-cyanoindene~\cite{sita2022}, 1- and 2-cyanonaphthalene~\cite{mcguire2021}, 1- and 5-cyanoacenaphthylene~\cite{cernicharo2024}, and 1-, 2-, and 4-cyanopyrene~\cite{wenzel_detection_2024,wenzel_detections_2025}. The only pure (unsubstituted) PAH indene was detected both in GOTHAM observations with the 100\,m Robert C. Byrd Green Bank Telescope~\cite{burkhardt2021} and in QUIJOTE observations using the Yebes 40\,m radio telescope~\cite{cernicharo2021}. Two possible gas-phase formation pathways, one of which occurs at low temperature, have only recently been uncovered~\cite{mccabe_off_2020,doddipatla_low-temperature_2021}. Many of the detected PAHs contain pentagonal structures, all based on cyclopentadiene~\cite{cernicharo2021}, including two isomers of cyanocyclopentadiene~\cite{mccarthy_interstellar_2021,lee2021} together with two ethylnylated cyclopentadiene isomers~\cite{cernicharo2021}, and the aforementioned indenes and cyanoacenaphthylenes.
This is remarkable, because these mark the first radio-astronomical detections of molecular species with pentagonal structures. 
On the other hand, the fullerene family of molecules (\ce{C60}, \ce{C60+}, and \ce{C70}) -- the largest molecules detected so far in space -- were found via optical and infrared measurements and contain (exclusively) pentagonal and hexagonal carbon rings \cite{cami_detection_2010,campbell_laboratory_2015}. 
The fully dehydrogenated versions of indene and fluorene, and also their charged variants, are direct building blocks of fullerenes.

Of the more than 300 molecules detected in the ISM to date, only 8 are anions and these have only been observed in the last 20 years~\cite{mcguire2022}. None of them are cyclic species, but successful detections demonstrate that large hydrocarbon anions containing as many as 10 carbon atoms are present~\cite{mccarthy_laboratory_2006,brunken_detection_2007,kentarou_observation_2007,cernicharo2023,millar2017,remijan2023}, and models predict that PAH anions are the dominant carriers of negative charge in cold and dense molecular clouds like \mbox{TMC-1}~\cite{wakelam2008}, where both the anions and PAHs mentioned above were detected. 
It is worth noting that the distinction between ``molecules’’ and ``grains’’ is generally a matter of convention, 
and grains as small as $3\;\text{\AA}$ (approximately the size of an indenyl molecule) are considered in studies of 
collisional charging of interstellar grains \cite{draine_collisional_1987}. 
In general, the charge states of PAHs in hotter and UV-irradiated environments such as photo-dissociation regions (PDRs) have been constrained to their neutral or cationic forms while their electron affinities ($E_\mathrm{ea}$) show that PAHs, and in particular dehydrogenated radical PAHs, are viable candidates for electron attachment~\cite{malloci2005,carelli_polycyclic_2012,carelli2013,buragohain2018}. This ability is enhanced when a pentagonal ring is present in the molecule~\cite{mishra2015}, making the study of these anions crucial to expanding our understanding of the physico-chemical processes at play in the ISM. 

\begin{figure}[t!]
  \centering
    \includegraphics[]{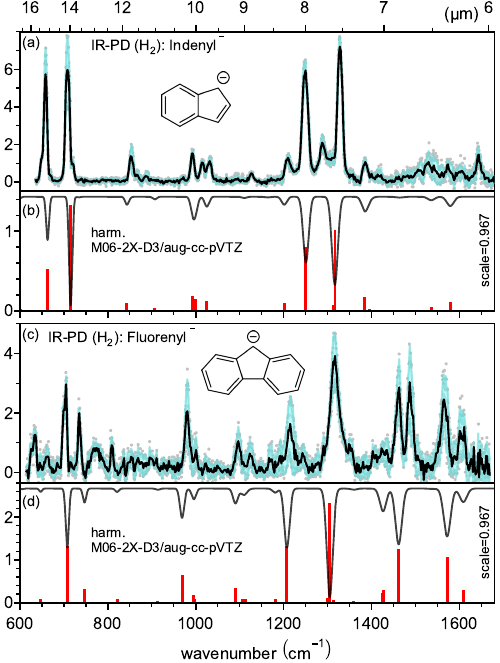}
    \caption{
    IRPD spectra (black) of the (a) indenyl and (c) fluorenyl anions tagged with \ce{H2} resulting from averaging all measurements for each wavenumber step (grey dots) with error envelope (cyan shading). The corresponding harmonic frequency stick spectra computed at the M06-2X-D3/aug-cc-pVTZ level of theory are depicted in red with inverse intensity (panels (b) and (d)). A scaling factor of 0.967 was applied. The inverted traces represent the computed spectra convolved with Gaussians with a bandwidth of $0.5\,\%$ of the corresponding wavenumber representative for the FELIX-2 laser line profile.
    }
    \label{fig:anions1}
\end{figure}

Here, we aim to investigate the pentagon-containing, planar PAH species indenyl (\ce{C9H7}, $115\;\massU$) and its related extension fluorenyl (\ce{C13H9}, $165\;\massU$) in both their anionic and cationic states using infrared pre-dissociation (IRPD) spectroscopy. Their molecular structures are depicted as insets in Fig.~\ref{fig:anions1}. The electron affinities for the indenyl and fluorenyl radicals have been previously determined to be $42.7\pm0.3\,\mathrm{kcal/mol}$ ($1.852\pm0.013\, \mathrm{eV}$) and $43.1\pm0.3\,\mathrm{kcal/mol}$  ($1.869\pm0.013\, \mathrm{eV}$), respectively~\cite{romer1997}, making them prone to electron attachment.

Indenyl and fluorenyl anions were previously spectroscopically examined by means of photoelectron velocity-map imaging spectroscopy~\cite{kim2013}; approximately 20 and 30 well-resolved vibronic transitions were identified from the anion and the radical neutral, respectively. An attempt to assign some of these features to theoretical spectra computed by density functional theory (DFT) delivered limited success, and no information on the vibrational modes of the anions was deduced. The anionic indenyl photo-detachment spectrum was later revisited by \citet{kumar2019}. Using \textit{ab initio} (MP2) calculations, they assigned the fundamental, combination, and overtone bands of the neutral radicals, and calculated values for the anion vibrational modes. Thus, to date there is no experimental data on the vibrational spectra of the indenyl and fluorenyl anions, 
and only limited direct vibrational gas-phase spectroscopy of radical \ce{[PAH]-} and 
deprotonated \ce{[PAH-H]-} anions in general: \citet{gao2014} studied 2-naphthyl, 9-anthracenyl, and 1-pyrenyl, using IR multiphoton electron detachment inside a Fourier transform ion-cyclotron resonance (FT-ICR) trap, 
and \citet{salzmann_infrared_2024} recorded the CH-stretching region of the Ar-tagged pyrene anion formed in a supersonic expansion.

It is interesting to note that, whereas the indenyl and fluorenyl anions are aromatic species, the cations do not fulfill H\"{u}ckel's rule for aromaticity because they contain $4n$
$\pi$-electrons, 8 and 12, respectively, and therefore are considered anti-aromatic. This is also supported by our computational nucleus-independent chemical shifts (NICS) analysis (see section S4 in the Supporting Information (SI)). 
Nevertheless, the stability of the fluorenyl cation has been demonstrated as it was identified as a common fragment in IR multiple photon dissociation (IRMPD) spectra when dissociating 9,10-dihydrophenanthrene, 9,10-dihydroanthrancene, and fluorene cations in ion trap experiments~\cite{petrignani_breakdown_2016}. The electronic spectrum recorded of the indenyl cation embedded in a solid Ne matrix revealed only four broad bands positioned up to $2000\,\rcm$ with tentative assignments using DFT calculations~\cite{nagy2013}. The fluorenyl cation has also been isolated in cold matrices. Its IR absorption spectrum has been recorded in an amorphous water ice matrix below $30\,\mathrm{K}$~\cite{costa2015}, while the electronic spectrum was measured in a Ne matrix at $6\,\mathrm{K}$, together with other isotopologues of \ce{C13H9+}
~\cite{fulara2016}. One of the most recent studies revealed a less perturbed IR spectrum of the ultra-cold fluorenyl cation and its two functionalized siblings in a He nanodroplet experiment ($\approx 10^5$ He atoms at $0.4\,\mathrm{K}$) using a cryogenic ($90\,\mathrm{K}$) hexapole ion trap accompanied by matching DFT calculations~\cite{greis2023}. Other recent studies concentrated on CN-functionalized hydrocarbons, namely the 2-cyanoindene \cite{douglas-walker_vibrational_2025} and cyanocyclopentadiene cations \cite{rap_spectroscopic_2022}. 

All these species mentioned here contain the pentagonal \ce{C5} ring as a not fully saturated hydrocarbon, such as cyclopentadiene. From a molecular physics perspective, the concept of (anti)aromaticity is pivotal in understanding the structure and stability of such polycyclic hydrocarbons, and introducing a charge to pentagon-containing PAHs will make a system (anti)aromatic with unique electronic properties. As for the above-mentioned cyclopentadiene, in its anionic charge state, it is considered to be aromatic, while its cationic counterpart is anti-aromatic~\cite{allen2001, abdellaoui2022}. We point out that two other cations with $4n+2$ $\pi$-electrons have recently been studied with IR and UV predissociation experiments and their vibrational and electronic spectra were recorded, $n=0$: \ce{c-C3H3+}\cite{marimuthu2020}, and $n=1$: \ce{c-C7H7+}\cite{jusko_direct_2018,jacovella_electronic_2020}.

Here, we use IRPD spectroscopy with a molecular hydrogen (\ce{H2}) tag to the ions of interest probed in the cryogenic 22 pole ion trap tandem mass spectrometer, FELion~\cite{jusko2019}, coupled to the widely tunable free electron lasers at the FELIX Laboratory (see Experimental Methods for details).\cite{oepts_free-electron-laser_1995}

The recorded IRPD spectra of the indenyl and fluroenyl anions tagged with \ce{H2} are presented in Fig.~\ref{fig:anions1}~(a) and (c), respectively. We compare them to predicted spectra of these anions determined using harmonic theoretical calculations at the M06-2X-D3/aug-cc-pVTZ level of theory~\cite{zhao_m06_2008,grimme_consistent_2010} that have been scaled with an empirical factor of 0.967 (factor for this method/basis set according to CCCBDB\footnote{\url{https://cccbdb.nist.gov/vsfx.asp}}) to obtain a good match to the experimental spectra; these calculated spectra are depicted in Fig.~\ref{fig:anions1}~(b) and (d). We can assign several fundamental bands by comparing their band centers 
to out-of-plane CH ($\gamma_\mathrm{CH}$), in-plane CH bending ($\delta_\mathrm{CH}$), and CC stretching ($\nu_\mathrm{CC}$) modes, or modes with mixed character of these as listed in Table~S3.1 in the SI. 

For the indenyl anion, the four strongest modes are at 658, 709, 1249, and 1328$\,\rcm$ which we can assign to $\gamma_\mathrm{CH}$ (658 and 709$\,\rcm$) and mixed character $\nu_\mathrm{CC}$ and $\delta_\mathrm{CH}$ (1249 and 1328$\,\rcm$) modes. Most of the weaker features in the IRPD spectrum of \ce{C9H7-} are accounted for in our theoretical harmonic IR frequencies, however,  features at 1015 and 1290\,$\rcm$, and those above 1400$\,\rcm$, do not have clear counterparts in the theoretically computed harmonic IR spectrum of the indenyl anion and therefore cannot be assigned. They might belong to combination bands of the indenyl anion or are shifted from the calculated values due to unaccounted anharmonic shifts. 
Although we cannot completely rule out the presence of another \ce{C9H7-} isomer in low abundance/concentration, 
none of the calculated spectra of selected isomers indicate that this would be the case (see Fig.~S5.1 in the SI).

The IRPD spectrum of \ce{C13H9-} can be assigned to the fluorenyl anion and most of the observed IRPD bands have a corresponding harmonic frequency counterpart. The two strongest modes of the fluorenyl anion at 703 and 1316$\,\rcm$ belong to the out-of-plane CH ($\gamma_\mathrm{CH}$) mode and a mode with mixed character of the CC stretch and in-plane CH bending motions ($\nu_\mathrm{CC}$ and $\delta_\mathrm{CH}$), respectively. In the range from 1400 to 1500$\,\rcm$, the assignment is not so clear. There are two strong bands in the IRPD spectrum of \ce{C13H9-} located at 1461 and 1487$\,\rcm$, while a much weaker feature appears at 1421$\,\rcm$. The latter might be assigned to the $\nu_\mathrm{CC} + \delta_\mathrm{CH}$ mode of the fluorenyl anion at 1424$\,\rcm$ and matching the 1461$\,\rcm$ with the 1462$\,\rcm$ mode. However, this would leave the second strongest IRPD feature at 1487$\,\rcm$ unassigned. We therefore propose also an alternative assignment, which is listed in parentheses in Table~S3.1 in the SI, considering the feature at 1421$\,\rcm$ to be in the noise of the experiment. In any case, the origin and assignment of these features are not completely clear, since we could also expect the overtone of the strong $\gamma_\mathrm{CH}$ out-of-plane CH bending mode in this region.

Notably, one of the strongest features in the IRPD spectra of both anionic species arises between 1250 and 1330$\,\rcm$ (8.0 to 7.5\,$\upmu$m) and correspond to CC stretching modes. As such, it is one of the most prominent PAH emission features centered around 7.7\,$\upmu$m, which plays a significant role in tracing PAH populations in the ISM that are influenced by the local radiation field and environmental conditions~\cite{bauschlicher_infrared_2009,stock_polycyclic_2017}. 
Just recently, detection of this band in unprecedented spatial resolution using the James Webb Space Telescope (JWST) uncovered star formation processes in the luminous infrared galaxy VV 114~\cite{evans_goals-jwst_2022}, consisting of the ongoing interaction or merger of two galaxies. Our measurements show that the indenyl and fluorenyl anions could contribute to this prominent PAH feature in regions with strong UV radiation fields.

\begin{figure}[t!]
  \centering
    \includegraphics[]{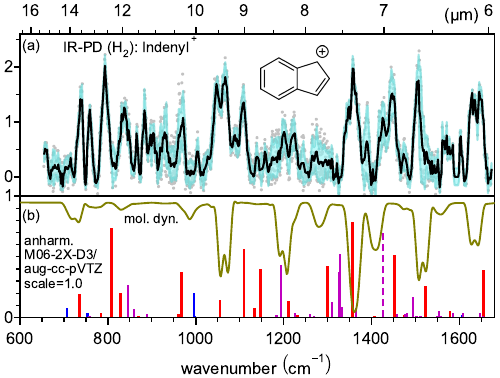}
    \caption{
    (a) IRPD spectrum (black) of the indenyl cation tagged with \ce{H2} resulting from averaging all measurements for each wavenumber step (grey dots) with error envelope (cyan shading).
    (b) The corresponding frequency stick spectrum computed at the anharmonic M06-2X-D3/aug-cc-pVTZ level of theory is depicted in red (fundamental bands), blue (first overtones), and, magenta (combination bands); theoretically calculated intensity is in $100\,\kJmol$ (note the unreliability for overtones/combination modes (dashed sticks), see text). No frequency scaling factor was applied. The inverted trace (olive) represents molecular dynamics simulation at the PBE-D3/DZVP level.}
    \label{fig:cations}
\end{figure}

For the indenyl cation, the spectral assignment is somewhat less conclusive based on scaled harmonic calculations alone, as shown in Figure S5.2 (b). In particular the predicted strong mode at 1150$\,\rcm$\ (mixed character CH and CC in plane bending) is not present in the experimental spectrum. Other isomeric structures explored theoretically at the same level of theory do not provide a conclusive match, either (see Figure S5.2 (c)--(h)). However, similarly to our previous studies on the cationic H-loss fragments of 2-methylanthracene~\cite{wenzel_infrared_2022} and aniline~\cite{rap_spectroscopic_2022}, or the cationic \ce{C2H2}-loss fragment of anthracene and phenanthrene~\cite{banhatti_formation_2022}, we cannot exclude the presence of several isomers in the experimental spectrum. We thus performed additional anharmonic calculations, as well as molecular dynamics simulations, which are shown together with the experimental IRPD spectrum in Figure \ref{fig:cations}. As can be seen, the molecular dynamics calculations capture most of the observed features, although the calculated and measured intensities mismatch in particular below 1000$\,\rcm$. Remarkably, the strong 1150$\,\rcm$ feature discussed above is weaker in intensity and now matches the observed experimental features, whereas on the other hand the experimental band at 1100$\,\rcm$ is now either shifted or is entirely absent. Similarly, a large deviation is seen around 1400--1500$\,\rcm$. The anharmonic DFT calculations predict the low frequency modes better, as well as the two experimental modes mentioned above, but fail to reproduce the experimental spectrum in other frequency regions. Within the time-independent calculations, most intense absorptions below 800$\,\rcm$ correspond to out-of-plane CH bending ($\gamma_\mathrm{CH}$), followed by in-plane CH bending ($\delta_\mathrm{CH}$) up to 1500$\,\rcm$, followed by CC stretching ($\nu_\mathrm{CC}$) above this wavenumber.

In conclusion, we recorded the IRPD spectra of three pentagon-containing PAH ions, namely \ce{C9H7-}, \ce{C13H9-}, and \ce{C9H7+}. In the case of the anionic species, we could assign our IRPD spectra by comparison to DFT calculated harmonic IR modes of indenyl and fluorenyl anions, respectively. To our knowledge, these are the first experimental gas-phase mid-IR spectra of these PAH anions, adding to the sparse existing experimental data on deprotonated PAH anions. We attempted to assign the IRPD spectrum of the cationic \ce{H}-loss fragment of indene, \ce{C9H7+}, to the indenyl cation. However, the IRPD spectrum was convoluted in a way that neither harmonic nor anharmonic DFT calculations at the M06-2X-D3/aug-cc-pVTZ level of theory could fully account for the observed IR features. We note that due to the perturbation treatment, the intensities of the overtones might not be correctly predicted. Molecular dynamics simulations were employed to compute a theoretical IR spectrum of the indenyl cation, which provides a more favorable match. The vibrational spectrum of the indenyl cation seems thus to be influenced by dynamic effects, making it unsuitable for static, time-independent DFT calculations, a method often used to interpret astronomical observations. The presence of other cationic isomers of \ce{C9H7+} can, however, not be excluded. With the identification of the indenyl and fluorenyl anions as contributors to the 7.7\,$\upmu$m PAH emission band, these experiments pave the way to further benchmark time-independent DFT calculations that can be used to analyze AIB spectra of PDRs and associated astronomical objects with high UV radiation fields, such as Titan's upper atmosphere in which negatively charged PAHs are also hypothesized to be present~\cite{zhao_low-temperature_2018,abplanalp_low-temperature_2019,lopez-puertas_large_2013,
nixon_composition_2024}. Especially with the large amount of new JWST data, further studies of PAH anions become necessary. In this context, we would like to highlight the recent advancements of the implantation of anions into He nanodroplet matrices, in which even closed-shell PAHs with low electron affinities $E_\mathrm{ea}\ll0.5\,\mathrm{eV}$ can form stable anions upon attachment of thermal electrons~\cite{gruber2022,kollotzek2022}. This might open a viable technique for future spectroscopic studies. 

\section{Experimental Methods}
\label{sec:exp}

The spectroscopic experiments were performed in the cryogenic radio-frequency 22 pole ion trap \cite{gerlich1992,asvany_coltrap_2014} located at the FELion beamline \cite{jusko2019} at the Free Electron Lasers for Infrared eXperiment (FELIX) Laboratory \cite{oepts_free-electron-laser_1995}. Indene (\ce{C9H8}) and fluorene (\ce{C13H10}) neutral samples (Sigma Aldrich, purity $\geq 99\,\%$ and $98\,\%$, respectively) were used without further purification. These precursors were vaporized by careful heating of the sample reservoir and directly fed into a storage ion source \cite{gerlich1992}. The indenyl and fluorenyl ions were produced by electron bombardment in dissociative ionization (indenyl cation) and in dissociative electron attachment (anions), delivering the positively or negatively charged \ce{H}-loss fragments of the precursor species. This procedure was first optimized in a similar 22 pole ion trap setup, Cold CAS Ion Trap (CCIT)\cite{jusko_cold_2024}, where an optimum electron energy, $E_{e^-}$, of approximately $50\,\text{eV}$ for anion production has been found. The produced ions were subsequently mass selected using a quadrupole mass filter and injected into the cryogenically cooled 22 pole ion trap held at approximately $10\,\text{K}$ to avoid \ce{H2} freeze out. High quantities of trapping He gas were pulsed into the trap at the same time as ion injection in order to a) trap the ions, and b) remove their internal and kinetic energy, \ie, cool the ions to temperatures close to that of the trap wall. After a set storage time, the ions were extracted and mass selected in a second quadrupole mass filter prior detection using an MCP counting detector. This experimental sequence, describing one cycle, was repeated with a typical period of $3\,\text{s}$. In order to switch between cation/anion operation of the setup only the relevant electrical potentials had to be adjusted.

In IRPD spectroscopy, the action scheme of dissociating a previously formed ion-tag complex upon resonant IR photon excitation is applied. Here, a 3:1 \ce{He}:\ce{H2} gas mixture was used as a trapping gas in order to promote ternary \ce{H2} attachments during the injection period. Therefore, the ion-tag complexes are efficiently produced only during the initial high number density injection period (approximately first $100\,\text{ms}$; total number density $>10^{15}\,\text{cm}^{-3}$), and no other reaction influences the number of ions inside the trap during the remaining storage time. We found that the He and Ne ternary sticking reaction rate was too low to attach these noble gases to any of the studied charged species in quantities sufficient for a reliable spectroscopic study. The widely tunable mid-IR light of the free electron laser, FELIX-2\footnote{\url{www.ru.nl/felix/}}, operated at $10\,\text{Hz}$ macropulse repetition rate, delivered on the order of $10\,\text{mJ}$ into the ion trap. The spectra, \ie, the number of the remaining ion-tag complexes at the end of the storage time and after laser irradiation (typically $2.6\;\text{s}$) inside the ion trap as a function of the laser wavenumber, are presented in Figs.~\ref{fig:anions1} and \ref{fig:cations}. Intensities were baseline corrected to take into account any background processes and normalized to the number of ions in the trap, the number of laser pulses, and, to the number of photons \cite{jusko2019,marimuthu2020}. The resulting signal is plotted as a positive quantity and the spectral resolution is only limited by the laser bandwidth. The laser wavelength was calibrated using a grating spectrometer, resulting in typical calibration uncertainties of $1-3\,\rcm$.

\section{Computational Methods}
\label{sec:comp}

Infrared spectra of the ions were calculated through both time-independent and time-dependent approaches. As for time-independent approaches, we used several density functionals (M06-2X, $\omega$B97XD, B3LYP) along with the aug-cc-pVTZ basis set, see the SI for benchmarking. Here, frequencies were calculated using both harmonic approximation and anharmonic calculations within second-order perturbation theory. In time-dependent calculations, we employed molecular dynamics at the temperature of 100\,K at the PBE/DZVP level with a time step of $0.5\,\mathrm{fs}$ and total running time of at least 100\,$\mathrm{ps}$. Along the trajectory, Wannier centers were calculated every five steps, subsequently used to produce the vibrational spectra. Aromaticity was analyzed using the nucleus-independent chemical shifts (NICS)~\cite{schleyer1996}. Time-independent calculations were performed in Gaussian \cite{frisch2016}, molecular dynamics in CP2K \cite{hutter2014}, analysis of the molecular dynamics in TRAVIS \cite{brehm2011,brehm2020}.

\begin{acknowledgement}


This work was supported by the Max Planck Society. 
The research leading to these results was supported by the funding received from LASERLAB-EUROPE, Grant Agreement no. 871124, under the European Union’s (EU) Horizon 2020 research and innovation programme. 
G.W. and B.A.M. acknowledge the support of the Arnold and Mabel Beckman Foundation 
Beckman Young Investigator Award. 
We gratefully acknowledge the support of Radboud University and of NWO for providing 
beam time at the FELIX Laboratory as well as the skillful assistance of the FELIX staff. 
We thank the Cologne Laboratory Astrophysics group for providing the FELion ion trap 
instrument for the current experiments and the Cologne Center for Terahertz Spectroscopy, 
funded by the Deutsche Forschungsgemeinschaft (DFG, grant SCHL 341/15–1), 
for supporting its operation.
The computational results presented have been achieved using the HPC infrastructure
LEO of the University of Innsbruck.

\end{acknowledgement}

\begin{suppinfo}


The following files are available free of charge.
\begin{itemize}
  \item Filename: SI.pdf: Calculated structural parameters, harmonic/ anharmonic frequencies, NICS, 
  harmonic frequencies for isomers of \ch{C9H7^{+/-}}.
\end{itemize}

The dataset associated with this work is available under \url{https://doi.org/10.5281/zenodo.10572237}.

\end{suppinfo}


\bibliography{references}

\end{document}